\documentclass[aps,pra,twocolumn,floats]{revtex4}

\usepackage{graphicx}
\usepackage{amsmath}
\usepackage{amssymb}
\usepackage{units}
\usepackage{ulem}
\usepackage{bm}
\usepackage{textcomp}

\begin{document}
\title{Generating long sequences of high-intensity femtosecond pulses}

\date{\today}
\author{M.~Bitter and V.~Milner}
\affiliation{Department of  Physics \& Astronomy and The
Laboratory for Advanced Spectroscopy and Imaging Research
(LASIR), The University of British Columbia, Vancouver, Canada \\}

\begin{abstract}{
We present an approach to create pulse sequences extending beyond 150~picoseconds in duration, comprised of $100~\mu$J femtosecond pulses. A quarter of the pulse train is produced by a high-resolution pulse shaper, which allows full controllability over the timing of each pulse. Two nested Michelson interferometers follow to quadruple the pulse number and the sequence duration. To boost the pulse energy, the long train is sent through a multi-pass Ti:Sapphire amplifier, followed by an external compressor. A periodic sequence of 84~pulses of 120~fs width and an average pulse energy of 107~$\mu$J, separated by 2~ps, is demonstrated as a proof of principle.
}
\end{abstract}

\maketitle

\section{Introduction}

Series of ultrashort laser pulses, also known as \textit{pulse trains} (PT), have found widespread use in the field of quantum coherent control of matter with laser light (for a recent review of this topic, see \cite{Voll2009} and references therein). Numerous applications require multiple pulses of relatively high intensity, on the order of $10^{13}$  to $10^{14}$~W/cm$^{2}$, to attain the regime of strong-field interaction with each pulse, just below the damage threshold of the material system under study.
Using coherent control of molecular rotation - an area of our own expertise - as only one representative example:
Sequences of intense ultrashort pulses have been key in enhancing molecular alignment~\cite{Bisgaard2004, Lee2004, Cryan2009} and aligning asymmetric top molecules in three dimensions~\cite{Lee2006, Ren2014},
selective excitation of molecular isotopes and spin isomers~\cite{Fleischer2006, Fleischer2007}, initiating uni-directional rotation~\cite{Fleischer2009, Kitano2009, Zhdanovich2011} and controlling gas hydrodynamics~\cite{Zahedpour2014}, as well as studying the quantum $\delta $-kicked rotor in a series of recent fundamental works~\cite{Zhdanovich2012, Floss2015, Kamalov2015, Bitter2015a}.
The great utility of PTs stems from two main factors. First, by matching the timing of pulses in the train to the dynamics of the system of interest, e.g. the vibrational or rotational period of a molecule, one can often significantly improve the selectivity of excitation. Second, the ability to re-distribute the energy among multiple pulses without losing the cumulative excitation strength enables one to avoid detrimental strong-field effects, such as molecular ionization and gas filamentation.

There are two common techniques to produce a pulse train with variable time separation between transform-limited pulses.
In the first technique, the incoming laser pulse is split into $2^n$ pulses using $n$ nested Michelson interferometers (MI). Even though sequences of up to 16~pulses~\cite{Siders1998} have been generated using this method, the scheme becomes increasingly more difficult to implement with the increasing value of $n$. We also note, that the control over the pulse timing is rather limited in that it cannot be changed independently for each individual pulse in the train. Similarly limited flexibility is characteristic of a pulse splitting method based on stacking a number of birefringent crystals~\cite{Zhou2007b}.

The second common approach is based on the technique of femtosecond pulse shaping (PS) where the spatially dispersed frequency components of the pulse are controlled in phase and amplitude~\cite{Wefers1993,Dugan1997}, or via the direct space-to-time conversion~\cite{Leaird1999}.
This offers much higher flexibility at the expense of being limited to the relatively low energy trains. The latter limitation is due to both the damage threshold of a typical pulse shaper, and also the necessity to block multiple spectral components in order to generate a train of pulses in the time domain.
Phase-only shaping has been often used to create a series of pulses without the loss of energy~\cite{Weiner1993}, but in this case, the distribution of the pulse amplitudes within the train is uneven and no control over this distribution is available.

Here we report on the generation of femtosecond pulse trains, that simultaneously satisfy the following specific characteristics: (1) consist of a large number of transform-limited pulses; (2) exhibit a relatively flat amplitude envelope; (3) can be easily tuned in terms of the timing of the constituent pulses; and (4) carry energies in excess of $100~\mu$J per pulse. Our method is based on the combination of a pulse shaper, which provides the often required flexibility in controlling the timing and amplitudes of individual pulses on the time scale of 50~ps, and a set of nested Michelson interferometers, which enables extending the overall length of the train to much longer times. Key to our approach is the integration of a multi-pass amplifier (MPA), which compensates the energy loss during the pulse shaping stage. We note that although amplification of shaped pulses is commonly used in chirped-pulse amplifiers~\cite{Efimov1998b,Pastirk2006}, and has also been employed to amplify pulse sequences~\cite{Liu1995,Dugan1997,Efimov1998a,Zhou2007}, to our knowledge, pulse trains with the above mentioned specifications have not been demonstrated.
We discuss the versatility of the developed tool and demonstrate an example of a long pulse train consisting of 84 equally strong pulses.

\section{Experimental setup} \label{Sec:Setup}

\begin{figure}
\centering
 \includegraphics[width=1.0\columnwidth]{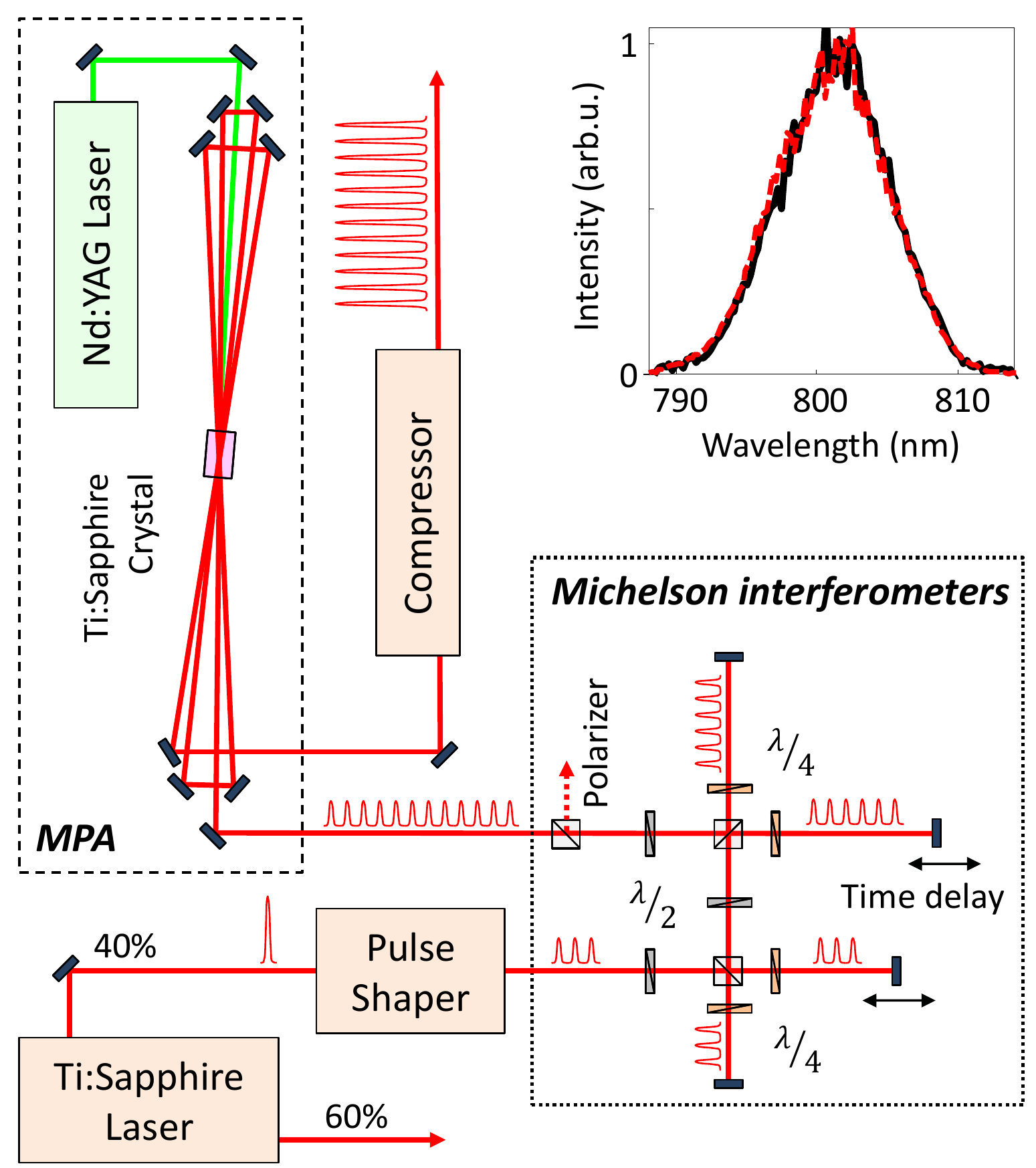}
     \caption{(color online) Optical setup for the generation of long sequences of high-energy femtosecond pulses. See text for details. Inset: Spectrum of a single pulse before (black solid, 0.4~$\mu$J) and after (red dashed, 100~$\mu$J) the amplification in the multi-pass amplifier (MPA). }
  \vskip -.1truein
  \label{Fig:Setup}
\end{figure}

We start with a Ti:Sapphire femtosecond laser system (SpitFire Pro, Spectra-Physics) producing uncompressed frequency-chirped pulses with the spectral bandwidth of 9~nm (full width at half maximum) at the central wavelength of 800~nm,  1~KHz repetition rate and 2~mJ per pulse. Part of the beam (60\% in energy) is compressed to 120~fs pulses (full width at half maximum) via a grating compressor and is used as a reference in cross-correlation measurements.
The second part is sent to a femtosecond pulse shaper which splits a single chirped pulse into a series of pulses, as schematically shown in Fig.\ref{Fig:Setup}. The shaper is built in the standard '$4f$'-geometry \cite{Weiner2000} and uses a liquid crystal spatial light modulator (SLM, Cambridge Research and Instrumentation, Inc.) with a double-layer 640-pixel mask.
We control the spectral phase and amplitude of the laser pulse to achieve the desired waveform of the output train. The required frequency masks are calculated through the Fourier transform of the target temporal profile (for more details see Ref.~\cite{Weiner2000}).
The spectral resolution of $\Delta \lambda=0.04\mathrm{nm}$ per pixel sets the upper limit for the total duration of the pulse train to $T=\lambda^2/{c\Delta \lambda} \approx 50\mathrm{ps}$, where the central wavelength $\lambda=800\mathrm{nm}$ and $c$ is the speed of light in vacuum.
The shaper is followed by two nested Michelson interferometers, shown inside the dotted box in Fig.\ref{Fig:Setup}. In contrast to the pulse shaper, the interferometric setup has no limitation on the overall length of the final pulse train. It allows to quadruple the number of pulses in the sequence and extend its duration to at least four times that produced by the shaper.

Unavoidable optical energy losses in the combined PS+MI setup can be quantified as follows.  About 50\% of energy is lost in a typical $4f$ pulse shaper due to the diffraction efficiency of the gratings. The energy throughput owing to the splitting of the initial pulse into a sequence of $N$ identical pulses by means of the spectral shaping can be estimated as $1/N$ \cite{Scaling}.
Another factor of 1/2 accounts for the polarization multiplexing in the MI setup. In total, each of the 4N~pulses in the train carries $1/(16N^2)$ of the available input energy. Since the latter is typically limited to about 1~mJ by the damage threshold of a pulse shaper (300~$\mu$J in our case), for a train of $4N=84$ pulses demonstrated in this work, one ends up with $\lesssim0.2$ $\mu $J per pulse. Imperfections of various optical components bring this number even further down, well below the typical requirements for a strong-field regime of the laser-molecule interaction.

To compensate for the energy losses described above, we send the long pulse train through a home-built multi-pass amplifier, schematically shown inside the dashed box in Fig.~\ref{Fig:Setup}. After passing four times through a Ti:Sapphire crystal, pumped by a Neodymium YAG laser (Powerlite Precision II, Continuum, 800~mJ at 532~nm and 10~Hz repetition rate), the pulses are amplified by a factor of up to 2800, reaching the energy of more than 100~$\mu$J per pulse. Both the pulse shaping and the interferometric splitting are applied to uncompressed frequency-chirped 150~ps pulses, so as to lower the peak intensities of the amplified pulses inside the MPA below the damage threshold of the Ti:Sapphire crystal. Chirped amplification also ensures negligible spectral distortions of the output pulses, as demonstrated in the inset of Fig.~\ref{Fig:Setup}. The spectrum of a single pulse amplified 250 times to an energy of 100~$\mu$J (red dashed line) is almost indistinguishable from the spectrum of an unamplified pulse at 0.4~$\mu$J (black solid line). The MPA is followed by a standard grating-based pulse compressor that compresses each pulse of the sequence to a 120~fs duration.

\section{Amplified Pulse Sequences}\label{Sec:AmplifiedPulseSequences}

\begin{figure}
\centering
 \includegraphics[width=1.0\columnwidth]{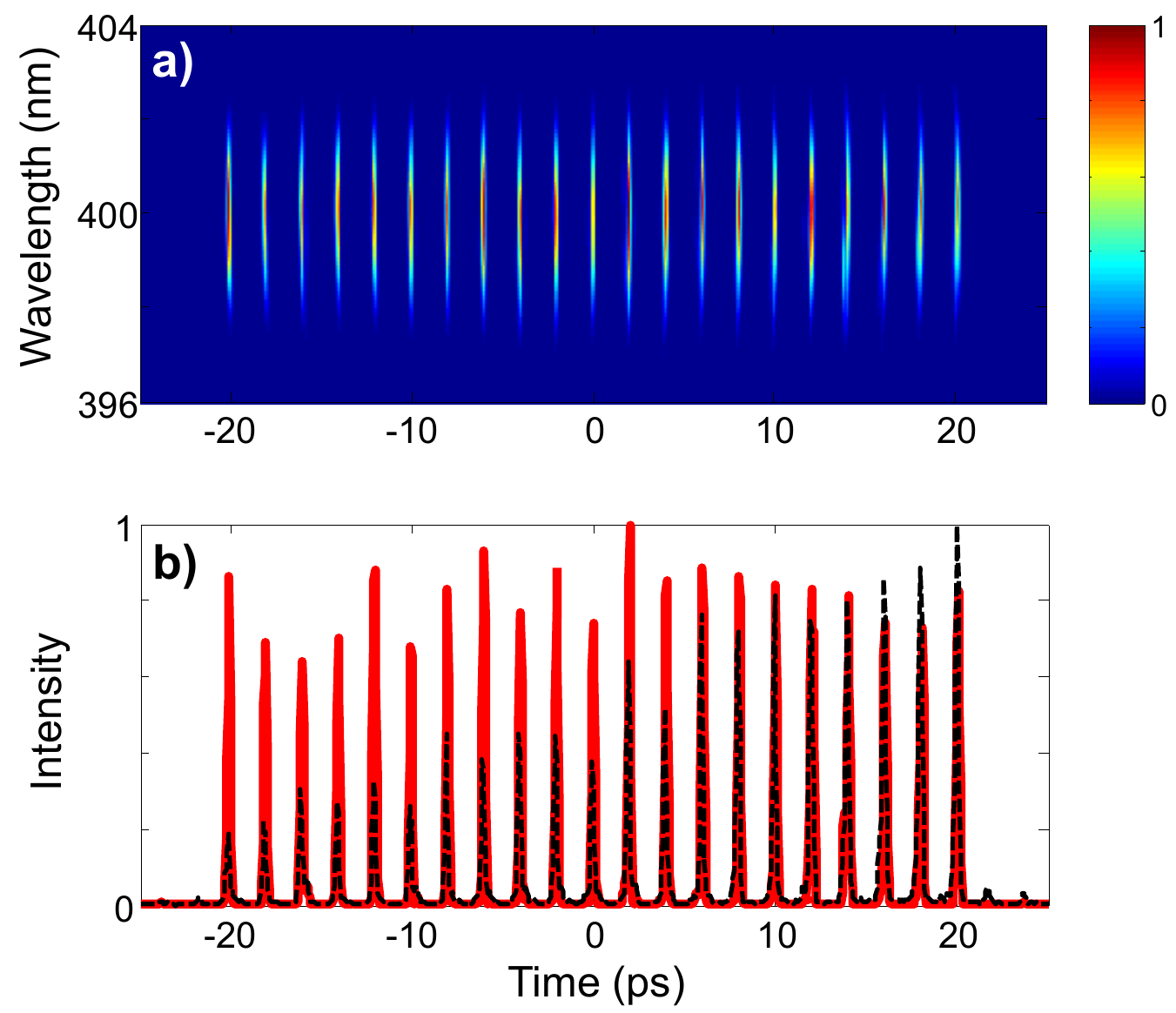}
     \caption{(color online) (\textbf{a}) xFROG spectrogram of an amplified pulse train with 21 equally spaced pulses and (\textbf{b}) its corresponding temporal profile before (dashed black) and after (solid red) the MPA amplification by a factor of 400.}
  \vskip -.1truein
  \label{Fig:PulseTrain_XFROG}
\end{figure}

First, we discuss pulse sequences which can be created with the pulse shaper only, i.e. without the use of Michelson interferometers. The pulse trains are characterized using the technique of cross-correlation frequency-resolved optical gating (xFROG) with a well-known transform-limited reference pulse. Figure~\ref{Fig:PulseTrain_XFROG}(\textbf{a}) shows an xFROG spectrogram for the pulse train of 21 equally spaced pulses with a period of 2~ps. The spectrogram demonstrates that each pulse in the train is transform-limited with no residual frequency chirp, confirming that proper pulse compression is attainable at the end of the combined shaping and amplification processes.

Integrating over the whole spectrum, one finds the distribution of energy among the pulses in the train, plotted in Fig.~\ref{Fig:PulseTrain_XFROG}(\textbf{b}). Here, the gain factor of 400 was required for reaching an average energy of 124~$\mu$J per pulse. Such high gain levels result in the increased sensitivity of the output pulse sequence to the energy fluctuations in the seed train entering the MPA, as well as in pulse dependent amplification rates. Equalizing the output amplitudes is achieved through the iterative process, in which the correction function is fed back to the pulse shaper in order to compensate for the irregularities in the output temporal profile. In the presented example, the corrected seed train is shown in Fig.~\ref{Fig:PulseTrain_XFROG}(\textbf{b}) by the black dashed line. The final degree of flatness is limited by the nonlinearity of the MPA amplification process and is on the order of $10\%$, determined as the standard deviation of the pulse-to-pulse energy fluctuations.

\begin{figure}[t]
\centering
 \includegraphics[width=1.0\columnwidth]{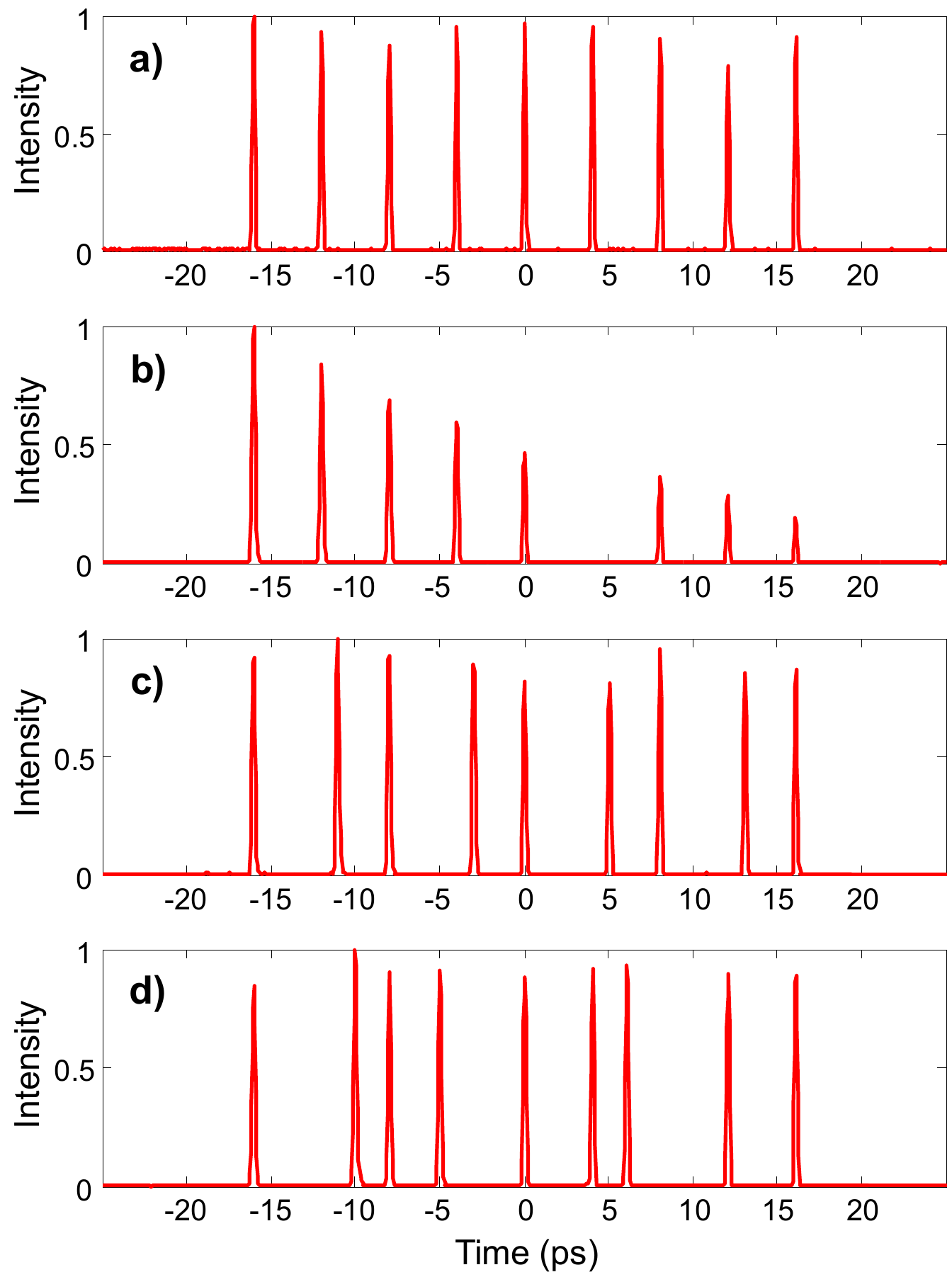}
     \caption{Temporal profiles of four different sequences of nine pulses:
     (\textbf{a}) Periodic train of equal-amplitude pulses separated by 4~ps.
     (\textbf{b}) Same train with linearly decreasing pulse amplitudes and the sixth pulse suppressed.
     (\textbf{c}) Flat pulse train with two different time periods of 5~ps and 3~ps.
     (\textbf{c}) Flat non-periodic pulse sequence with a random timing of pulses.}
  \vskip -.1truein
  \label{Fig:PulseTrain_versatile}
\end{figure}

Femtosecond pulse shaping offers the flexibility of creating arbitrary pulse sequences within the limits of the shaper's temporal and spectral resolution.
In Fig.~\ref{Fig:PulseTrain_versatile}, we demonstrate this flexibility using the example of a pulse train with nine pulses. The overall energy of the train was set to 1~mJ. A flat train of pulses with almost equal amplitudes ($7\%$ flatness) separated by $T=4\mathrm{ps}$ is shown in Fig.~\ref{Fig:PulseTrain_versatile}(\textbf{a}).
In Fig.~\ref{Fig:PulseTrain_versatile}(\textbf{b}), a linear amplitude tilt was applied to the train's envelope and its sixth pulse was completely suppressed, while the total energy was kept constant at 1~mJ.
Figures~\ref{Fig:PulseTrain_versatile}(\textbf{c}) and (\textbf{d}) demonstrate our ability to produce high-energy flat-amplitude sequences with multiple periods and completely random timing of pulses, respectively.

\begin{figure}
\centering
 \includegraphics[width=1.0\columnwidth]{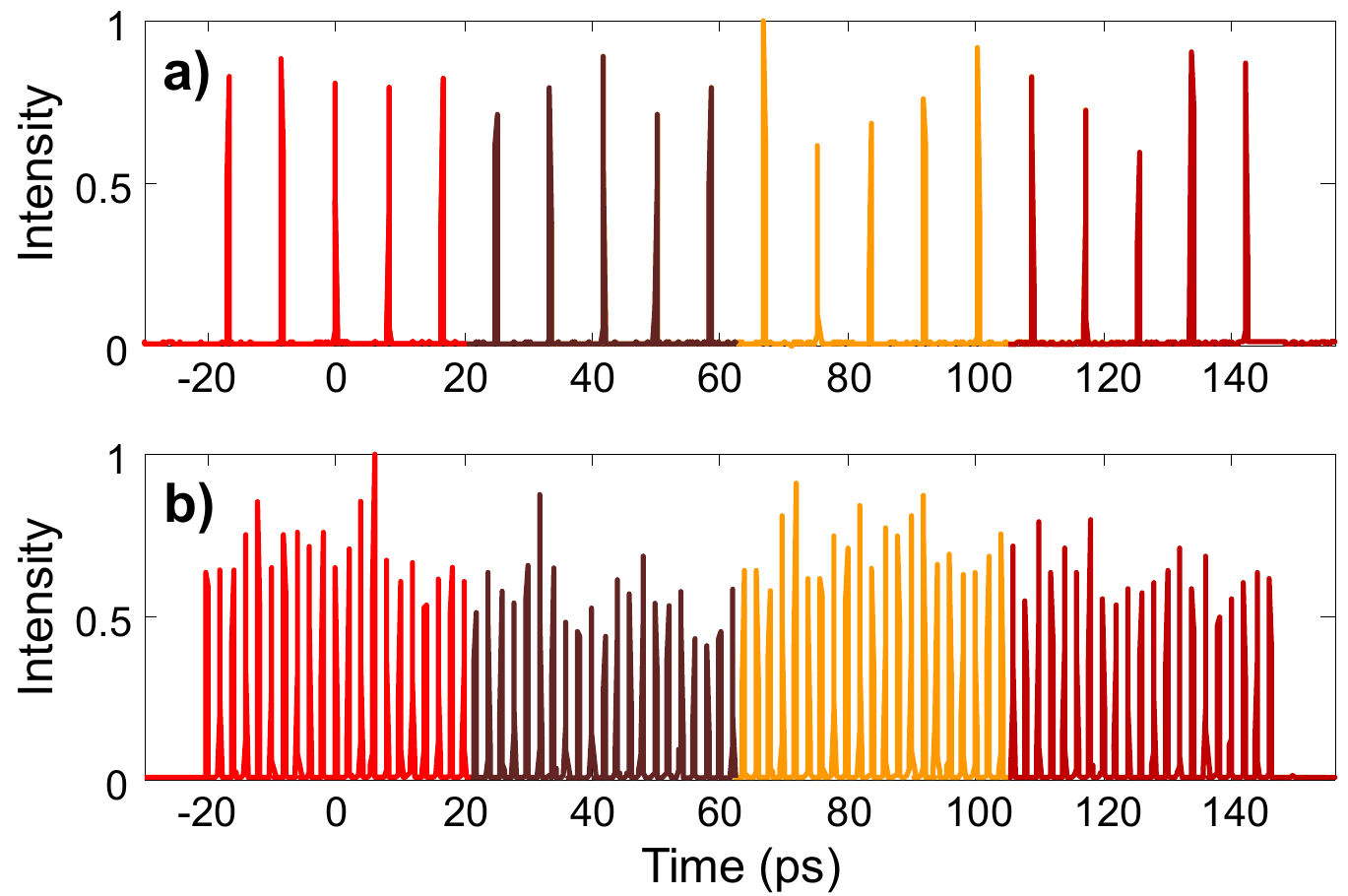}
     \caption{ (color online) Temporal profile of two periodic pulse
sequences:
(\textbf{a}) 20~pulses separated by 8~ps and amplified 500 times to 114~$\mu$J per pulse;
(\textbf{b}) 84~pulses separated by 2~ps and amplified 2800 times to 107~$\mu$J per pulse.
The four colors represent the four different pathways through the Michelson interferometers.}
  \vskip -.1truein
  \label{Fig:PulseTrain_long}
\end{figure}

Adding a set of nested Michelson interferometers enables us to generate pulse sequences longer than the time limit set by the spectral resolution of the shaper (in our case, 50~ps) and with more pulses, while still maintaining the energy level in excess of 100~$\mu $J per pulse. In Figures~\ref{Fig:PulseTrain_long}(\textbf{a}) and (\textbf{b}), we show periodic sequences of 20 and 84 pulses stretching over a duration of $\approx 170$~ps (pulse separation of 8~ps and 2~ps, respectively) and carrying $\approx110 \mu$J per pulse. The latter train has been amplified 2800 times to the total energy of 9~mJ. One can see that longer pulse trains suffer from a higher amplitude noise, e.g. standard deviations of $13\%$ and $18\%$ in Fig.\ref{Fig:PulseTrain_long}(\textbf{a}) and (\textbf{b}), respectively. The increasing noise is due to the higher MPA amplification factors and the correspondingly higher nonlinearity of the amplification process.

\section{Summary}\label{Sec:Conclusion}
We implemented an optical setup that produces long sequences of femtosecond pulses. The setup is based on the combination of femtosecond pulse shaping, interferometric pulse splitting and multi-pass amplification. The shaping offers the flexibility of creating arbitrary pulse trains on the time scale of 50~ps. The duration of the train can be further increased (in this work, fourfold) by means of the setup of nested Michelson interferometers. Finally, multi-pass amplification compensates for the severe energy losses typically associated with the generation of long pulse sequences from a single femtosecond pulse. An example of a train of 84~pulses separated by 2~ps with an energy exceeding 100~$\mu$J per 120~fs pulse is presented.

We are grateful to Jonathan Morrison and Kamil Krawczyk for their help with the optical setup.



\begin{thebibliography}{10}
\bibitem{Scaling}
The $1/N$ scaling can be derived from the following consideration of a pulse train in the frequency domain. The spectrum of an infinite pulse train is a frequency comb. The separation of the comb teeth is inversely proportional to the train period $T$, $\Delta \omega \propto 1/T$. The spectrum of a finite train of $N$ pulses is the same comb convolved with the sinc function whose width is inversely proportional to $NT$, $\delta \omega \propto 1/(NT)$. The ratio $\delta \omega / \Delta \omega $ determines the amount of energy remaining in the train after shaping.
\end{thebibliography}

\end{document}